# Phonons and phase symmetries in bulk CrCl$_3$ from scattering measurements and theory

X. Li[1], S.-H. Do[1], J.-Q. Yan[1], M. A. McGuire[1], G. E. Granroth[2], S. Mu[3], T. Berlijn[4,5], V. R. Cooper[1], A. D. Christianson[1*], and L. Lindsay[1†]

[1]Materials Science and Technology Division, Oak Ridge National Laboratory, Oak Ridge, Tennessee 37831, USA

[2]Neutron Scattering Division, Oak Ridge National Laboratory, Oak Ridge, Tennessee 37831, USA

[3]Materials Department, University of California, Santa Barbara, California 93106, USA

[4]Center for Nanophase Materials Sciences, Oak Ridge National Laboratory, Oak Ridge, Tennessee 37831, USA

[5]Computational Sciences and Engineering Division, Oak Ridge National Laboratory, Oak Ridge, Tennessee 37831, USA

Corresponding email: christiansad@ornl.gov* (experiment); lindsaylr@ornl.gov† (theory)

*Abstract*

Phonon-derived behaviors are important indicators of novel phenomena in transition metal trihalides, including spin liquid behavior, two-dimensional magnetism, and spin-lattice coupling. However, phonons and their dependence on spin structure and excitations have not been adequately explored. In this work, we probe and critically examine the vibrational properties of bulk $CrCl_3$ using inelastic neutron scattering and density functional theory. We demonstrate that magnetic and van der Waals interactions are essential to describing the structure and phonons in $CrCl_3$; however, the specific spin configuration is unimportant. This provides context for understanding thermal transport measurements as governed by dynamical spin-lattice couplings. More importantly, we introduce an efficient dynamic method that exploits translational symmetries in large conventional unit cells that generates insights into phonon dispersions, interactions, and measured spectra in terms of quantum phase interference conditions. This work opens new avenues for understanding phonons in layered magnets and more generally in conventional cell geometries of a variety of materials.

## 1. Introduction

The recent realization of two-dimensional (2D) magnetic systems has propelled intense research into the exploration of their fundamental properties and potential for practical applications[1-3]. A prominent material class for achieving 2D and quasi-2D magnetism is van der Waals (vdW) layered magnetic materials as they can be easily cleaved to a few-layer or monolayer form where the physical properties (*e.g.*, band gaps) show significant differences from their bulk counterparts[4-8]. Chromium trihalides $CrX_3$, where $X$ = Cl, Br, I are particularly interesting cleavable magnets with honeycomb layers and in-plane ferromagnetic (FM) ordering below their respective critical temperatures. These materials have potential for controlling spin and lattice degrees of freedom for applications including topological magnonics[9-11] and spintronics[12]. $CrCl_3$, in particular, has low temperature antiferromagnetic (AFM) interlayer interactions and is a promising material for studying spin and vibrational transport in magnetic field[13].

$CrCl_3$ is also a structural analog of $\alpha$-$RuCl_3$, a Kitaev quantum spin liquid candidate, with signatures of this state observed half-integer via the quantum thermal Hall effect[14-16]. Measurements of thermal conductivity in both systems (derived from phonons as these are insulating materials) have demonstrated large magnetic-field-induced enhancements at low temperatures, though above their respective Néel temperatures[17,18]. For both $\alpha$-$RuCl_3$[17] and $CrCl_3$[13], models based on suppression of spin-phonon interactions modulated by magnetic field were used to understand this peculiar thermal transport behavior. These models were built from simple isotropic Debye models and transport determined within a Debye-Callaway formalism[19,20] with empirical scattering terms based on a variety of fitting parameters to describe intrinsic phonon scattering and magnetic scattering, among others.

While satisfactory agreement between measurements and calculations was achieved for temperature- and magnetic-field-dependent thermal conductivities in $\alpha$-$RuCl_3$ and $CrCl_3$, significant misrepresentation of independent scattering mechanisms can occur in fitting procedures with many empirical parameters, particularly while using toy models of the phonon dispersions and transport. A more accurate description of the phonon dispersions, particularly their strong anisotropy and large number of optic bands, will provide better interpretation of the complex physics behind the measured data. Motivated by Raman measurements[21], phonon dispersions of bulk $CrCl_3$ have been recently calculated via density functional theory (DFT). However, no insights were developed regarding the dependence of vibrational mode behaviors under different conditions, including variations with magnetic configuration. Furthermore, there is a conspicuous lack of measured phonon dispersion data for bulk CrCl$_3$ in the literature beyond Raman-active optic modes[22-25], presumably due to synthesis challenges.

Here we present inelastic neutron scattering (INS) measurements and DFT-based calculations of the phonon spectra of bulk CrCl$_3$. First, we build a dynamical formalism based on translational symmetry within large conventional unit cells, such as in bulk CrCl$_3$ (see Fig. 1), in order to simplify calculations and derive insights into the vibrational phase relationships of complex layered systems. Using this, we show that the INS spectra can be understood in terms of quantum phase interference conditions and demonstrate that spin-resolved DFT-based calculations give a good description of the measured phonons of CrCl$_3$, comparing with INS data here and with previously measured Raman data. Finally, we justify the magnetic structures used in calculations and examine the effects of varying spin configurations, Hubbard corrections, vdW interactions, and spin-orbit couplings (SOC) on CrCl$_3$ phonons. Our study provides a detailed description of the vibrational properties of cleavable magnetic CrCl$_3$ and introduces a general dynamical formalism for understanding phonons and other quasiparticles in larger, more convenient conventional cells rather than abstract primitive cell descriptions where lattice vectors form relatively arbitrary angles in Cartesian space.

## 2. Theory
### 2.1 Atomic and magnetic structures

CrCl$_3$ is an anisotropic honeycomb vdW layered material consisting of hexagonal layers of Cr atoms with octahedrally coordinated Cl atoms. At temperatures below 240 K, it has rhombohedral structure with space group $R\bar{3}$ and ABC stacking defined by a translational vector **S**=[2/3, 1/3, 1/3] in fractional coordinates that describes the structural relationship of adjacent layers[26], see VESTA[27] visualization in Fig. 1. Below the Néel temperature ($T_N$=14 K) the Cr magnetic moments are aligned ferromagnetically in the plane and AFM across the plane[5,26], while above $T_N$ CrCl$_3$ is paramagnetic (PM). In this work, we examine the structure and phonon dispersions for a variety of magnetic states: non-magnetic (NM), low temperature AFM state, an in-plane AFM state, and a fully FM case for which all the spins are aligned, in-plane and cross-plane (see **Table 1** for list of calculations). The PM state was not explicitly examined as it requires significant computational resources to configurationally average large disordered supercells. We demonstrate later that the CrCl$_3$ phonons do not depend strongly on the spin configuration, as long as one exists. For the FM configuration, the conventional magnetic and structural cells are identical and can be described by three layers (2 Cr and 6 Cl atoms per layer). However, for the AFM structure, each magnetic unit cell consists of two structural unit cells thus requiring six layers in the conventional cell. We note that the interlayer interactions are weak[28] and the AFM structure can be switched to FM with a small in-plane magnetic field ~0.2 T[26,29]. We use the three-layer FM structure for most calculations in this work; however, we use a six-layer unit cell for AFM calculations and for the FM structure when comparing with AFM calculations.

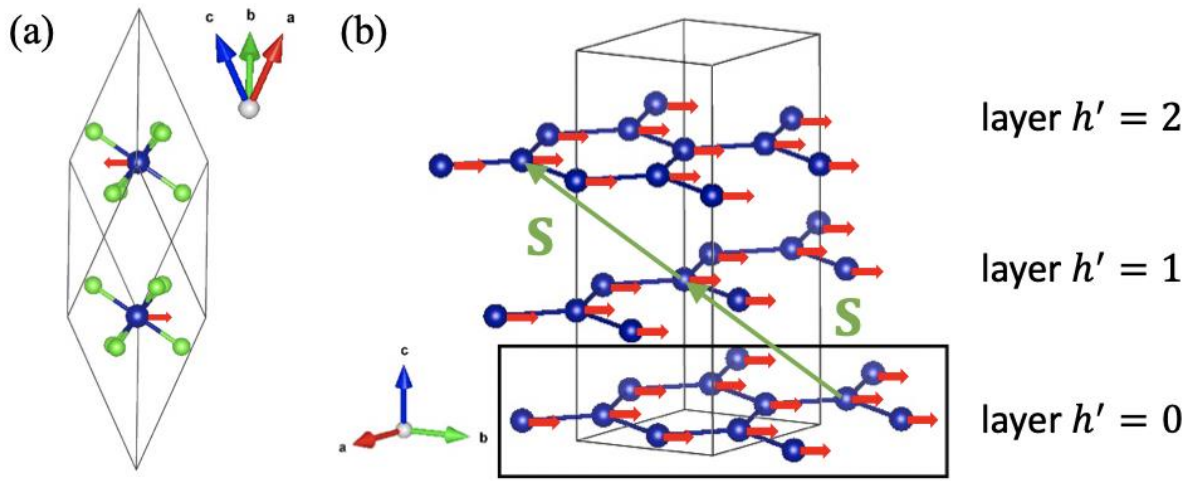

**Figure 1 | Crystal structure of CrCl$_3$ ($R\bar{3}$ space group).** (a) AFM configuration in a primitive unit cell with 8 atoms. (b) FM configuration showing the three-layer configuration (only Cr atoms are shown). The green arrows represent the translation vector **S**=[2/3, 1/3, 1/3] in fractional coordinates relating layers by translational symmetry. The black rectangle highlights that the translation-symmetry-based dynamics (see text) can be determined by a single layer, while other layer vibrations are phase related. The red arrows in (a) and (b) represent in-plane magnetic moments on Cr atoms.

Different magnetic configurations, Coulomb correlations, vdW corrections, and SOC were considered (summarized in **Table 1**). For the AFM interlayer configuration, the interatomic force constants (IFCs) were calculated based on a 192-atom 2x2x1 supercell. For the FM interlayer configuration, the IFCs were calculated for a 2x2x1 supercell with unit cells considering separately three and six layers, depending on the calculation (see **Table 1**). The DFT+$U$ method[30] was used to include the Coulomb correlations with $U_{\text{eff}}$=3.4 eV for Cr atoms, which is an average value from previous studies[31-34], and compared with calculations without this correction. Van der Waals interactions were taken into account via independent DFT-D3[35] and vdW-DF-optB86b[36-39] methods. The spin-orbit coupling was also included for some calculations to test its effect on the phonons.

**Table 1 | Summary of DFT calculations with different magnetic configurations, Hubbard $U$ corrections, vdW interactions, and SOC.** Measured $R\bar{3}$ lattice parameters at T=225 K are $a$=5.942 Å and $c$=17.333 Å[40]. Calculated $c$ parameters with the asterisk correspond to values for a 3-layer system. These need to be multiplied by two for the actual $c$ value for the 6-layer system. All calculations include the Coulomb correlations with $U_{\text{eff}}$=3.4 eV except for calculation (4) marked with the dollar sign.

| configuration | layer # | vdW | SOC | $a, c$ [Å] | appearance |
|---|---|---|---|---|---|
| (1) FM | 3 | DFT-D3 | Yes | 6.050, 17.391 | Figs. 2, 3, 6, S2, S3 |
| (2) FM | 6 | DFT-D3 | No | 6.052, 17.387* | Fig. 4 |
| (3) AFM | 6 | DFT-D3 | No | 6.052, 17.384* | Figs. 4, 5 |
| (4) AFM$ | 6 | DFT-D3 | No | 5.989, 17.253* | Fig. 4 |
| (5) FM | 3 | vdW-DF-optB86b | No | 6.006, 17.244 | Fig. 6 |
| (6) FM | 3 | DFT-D3 | No | 6.050, 17.394 | Figs. 5, 6, S4 |
| (7) FM | 3 | - | No | 6.122, 19.163 | - |
| (8) FM | 3 | - | Yes | 6.126, 19.436 | - |
| (9) NM | 3 | vdW-DF-optB86b | No | 5.888, 16.910 | - |
| (10) AFM | 3 | DFT-D3 | No | 6.037, 17.458 | Fig. S4 |

Regardless of magnetic state, the primitive unit cell of the underlying CrCl$_3$ lattice (Fig. 1a) does not provide an intuitive description of its structure. Thus, a larger conventional cell is most often used. In the next section we explore the relation between dynamical phases of different layers within the conventional cell description that sheds light on measured observables and quasiparticle interactions. The dynamical methods presented here are applicable to the conventional cells of a wide variety of layered and covalently bonded materials having internal translational symmetry (e.g., **S** in Fig. 1b) neither parallel nor perpendicular to the conventional lattice vectors, that relates the conventional cell to the primitive cell. We designate this mapping of atoms within the conventional cell 'primitive translational symmetry (PTS)' to highlight this relationship.

## 2.2 Primitive translational symmetry (PTS) in conventional cells

Phonon dispersions for conventional cells can be calculated from primitive unit cells with appropriate phase relations based on translational symmetries. Exploiting the PTS, the vibrational properties of a conventional cell can be determined with smaller computational cost and gives information of phase interference that acts as conservation conditions in the conventional geometry, which are useful for interpretating observed INS data (see Fig. 2) and in transport calculations.

We start from the traditional dynamical matrix that determines phonon dispersion relations:

$$D_{\alpha\beta}^{KK'}(\mathbf{q}) = \frac{1}{\sqrt{m_K m_{K'}}} \sum_{p'} \Phi_{\alpha\beta}^{0K,p'K'} e^{i\mathbf{q}\cdot\mathbf{R}_{p'}} \quad (1)$$

Using a conventional cell here, $m_K$ is the mass of the $K^{th}$ atom, Greek subscripts are Cartesian directions, $\Phi_{\alpha\beta}^{0K,p'K'}$ are harmonic IFCs between atom $K$ in the origin unit cell and atom $K'$ in the $p'$ unit cell, $\mathbf{q}$ is a wavevector, and $\mathbf{R}_{p'}$ is a lattice vector locating the $p'$ unit cell. For bulk CrCl$_3$ in the FM configuration, the conventional cell has 24 atoms (8 in each of three layers), and the dynamical matrix is a 72x72 matrix to be diagonalized for every $\mathbf{q}$.

Exploiting PTS within the conventional cell, the dynamics of conventional cells can be divided into smaller subunits (single layers / building blocks of the primitive unit cells). Unit cell atoms in a single layer become the new basis of the subunit, which are related to the other layers via fractional translation defined by the translation vector $\mathbf{S}$ (for bulk CrCl$_3$ with $R\bar{3}$ symmetry $\mathbf{S}$=[2/3, 1/3, 1/3] in fractional coordinates of a single conventional cell). The dynamical matrix with this new basis becomes:

$$D_{\alpha\beta}^{kk'}(\mathbf{q}, l) = \frac{1}{\sqrt{m_k m_{k'}}} \sum_{h'p'} \Phi_{\alpha\beta}^{0k,h'p'k'} e^{i\mathbf{q}\cdot(\mathbf{R}_{p'}+h'\mathbf{S})} e^{ilh'2\pi/N} \quad (2)$$

where $k$ loops over the atoms in a single layer, $h'$ is a layer index (illustrated in Fig. 1b), and $e^{ilh'2\pi/N}$ is the vibrational phase between layers with $N$ the number of layers in the conventional cell. In this formalism, the phonon wavevector has four components: a continuous

3D translational wavevector $\mathbf{q}=(q_x, q_y, q_z)$ and an integer $l$ representing the layer degrees of freedom within the conventional cell. The $l$ ranges from 0 to $N$-1 and importantly, is a conserved quantity in scattering processes, like the translational momentum $\mathbf{q}$. With this formalism, the conventional dynamical matrix (72×72 for FM bulk $CrCl_3$) is broken into smaller dynamical matrices (three 24×24 matrices) for each $l$, which reduces the computational cost of diagonalization, particularly for large systems. We apply now PTS in the conventional dynamics of bulk $CrCl_3$ and compare with INS measurements in the next section.

## 3. Experiment

For the inelastic neutron scattering measurements, the same single crystal of $CrCl_3$ as used in Ref. 28 was aligned on an aluminum plate with an $[H, K, 0]$ horizontal scattering plane. The INS data were measured at 30 K using the ARCS time-of-flight spectrometer[41] at the Spallation Neutron Source located at Oak Ridge National Laboratory. The Fermi chopper was phased for the high-resolution mode with setting to 300 Hz for an incident energy $E_i$=25 meV, which gives a FWHM=0.7 meV of resolution at the elastic line (0 meV). For $E_i$=25 meV, measurements were performed by rotating the sample through 212° about its vertical axis with 0.5° steps. The data were symmetrized using symmetry operations of the Laue class of the $R\bar{3}$ to enhance statistics.

## 4. Results
### 4.1 Phonon dispersions from INS and DFT

Figure 2 gives measured INS dynamical structure factor data for bulk $CrCl_3$ along a high-symmetry path at T=30 K (PM state) and compared directly with DFT calculations of the phonon dispersion for the FM configuration. We demonstrate later that the calculated dispersion is insensitive to varying spin configurations. We also calculate Raman-active phonon frequencies and compare with measurements[42] in **Table 2**.

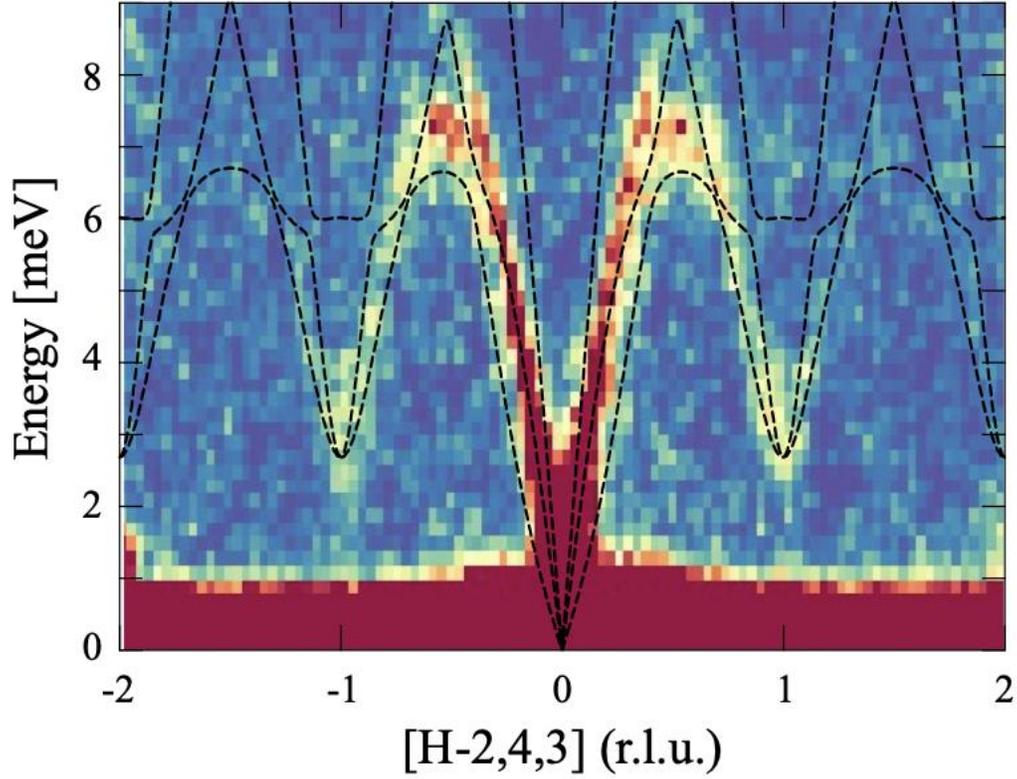

**Figure 2 | In-plane phonon dispersion of bulk CrCl$_3$.** INS measured dynamical structure factor (heat map) is compared with DFT-based phonon dispersion (dashed curves). Only dashed black curves with integer $l$=0 are accessible to INS measurements and shown in the figure. Measured dynamical structure factors without the theoretical overlays are given in the Supplemental Materials.

**Table 2 | Phonon frequencies of room temperature Raman-active modes.** DFT calculated frequencies compare favorably with Raman measurements[42].

| Points | $E_g$ | | | | $A_g$ | | | |
|---|---|---|---|---|---|---|---|---|
| Calculated frequency [meV] | 13.67 | 24.78 | 28.37 | 41.00 | 19.72 | 35.76 | 19.22 | 34.90 |
| Measured frequency [meV] | 14.45 | 25.86 | 30.56 | 42.72 | 20.50 | 37.23 | - | - |

The PTS that governs phase relations in the dynamics of conventional cells (Eq. 2) underlies interference conditions that can give insights into INS spectra of materials using conventional geometries. Using PTS, the conventional cell INS dynamical structure factor is[43-45]

$$S(\mathbf{Q}, E) = \frac{1}{2N_{uc}N} \sum_{j,l} \left| \sum_{h=0}^{N-1} e^{i2\pi lh/N} \sum_{k} \frac{b_k}{\sqrt{m_k}} e^{-i\mathbf{Q}\cdot\Delta_k}(\mathbf{Q} \cdot \boldsymbol{\varepsilon}_{\mathbf{Q},j,l,k}) e^{-W_{h,k}} \right|^2 \frac{(n_{\mathbf{Q},j} + 1)}{\omega_{\mathbf{Q},j}} \quad (3)$$

where $\mathbf{Q}$ and $E$ are momentum and energy transfers, respectively, $N_{uc}$ is the number of conventional cells in the crystal, $\omega_{\mathbf{q},j}$ is phonon frequency for wavevector $\mathbf{q}$ ($\mathbf{q} = \mathbf{Q}$ after application of momentum conservation) and polarization $j$, $n_{\mathbf{q},j}$ is the Bose-Einstein distribution, $e^{-W_{h,k}}$ is a Debye-Waller factor, $b_k$ is the average nuclear scattering length of atom $k$, $\Delta_k$ is the position of atom $k$ with respect to its layer, $\boldsymbol{\varepsilon}$ is the eigenvector determined by diagonalization of Eq. 2. See Supplemental Materials[46] (SM) for derivation of Eq. 3 from the conventional formalism. The important aspect to focus on here is the phase factor in the sum over layers. Summing over $h$ (first sum in brackets of Eq. 3) for $l=0$ gives:

$$l = 0: \sum_{h=0}^{N-1} e^{i2\pi lh/N} = \sum_{h=0}^{N-1} e^0 = N \quad (4)$$

while for $l \neq 0$ we can use the exponential sum formula:

$$\sum_{h=0}^{N-1} e^{ih(2\pi l/N)} = \frac{1 - e^{iN(2\pi l/N)}}{1 - e^{i(2\pi l/N)}} = 0 \quad (5)$$

since $e^{i2\pi l} = 1$ regardless of the value of $l$. This demonstrates that the dynamical structure factor vanishes for all phonons unless $l = 0$, suggesting that only these phonons can be observed by INS in the conventional geometry. This is confirmed in Figs. 2 and 3 for bulk CrCl$_3$ where the INS spectral intensity only appears with black $l=0$ calculated dispersions. For example, acoustic modes at [-3, 4, 3] and [-1, 4, 3] in Fig. 2 are not seen because they have $l \neq 0$. Note that this phase cancelation occurs for all $\mathbf{Q}$ in reciprocal space, unlike similar dynamical structure factor calculations in complex twisted materials that were defined only along one-dimensional high symmetry lines[45]. Such cancellation of the dynamical structure factor is known for unit cells having multiple formula units[45,47,48], but is explicitly shown here for specific phonon branches regarding translational symmetry in the unit cell.

Measured and calculated phonon dispersions along other high-symmetry directions are given in Fig. 3. General agreement is observed in all directions, including the high-frequency optical modes and separately measured Raman-active modes[42]. There is a slight softening of the theoretical frequencies with respect to measurements, typical of PBE calculations which tend to give larger lattice parameters, as found for bulk CrCl$_3$ here. Again, the observed phonons in INS and Raman measurements only sample branches with integer $l = 0$ in Eq. 2 using PTS in the conventional cell. Not all phonons with $l = 0$ are observed in the experiments, likely a result of suppressed intensity due to the dot product of the momentum transfer and eigenvectors for the differing probe geometries. In Fig. 4 we show extracted discrete measured data points compared with calculations along the $A - \Gamma - M - K - \Gamma$ high-symmetry line in the first Brillouin zone (fBZ). The experimental dispersions were extracted by a Gaussian fit to constant momentum scans, as shown in Fig. (S2) in the SM[46]. Note that the measured data here are mapped from varying zones and are thus not subject to the phase cancellation rules described above. Calculations demonstrate that the dispersion using PTS in Eq. 2 (dashed colored curves in Fig. 4) are indistinguishable from that of the conventional dynamics from Eq. 1 (solid green curves in Fig. 4). We also show the INS measurements without theoretical overlays and calculated phonons for the full frequency range of the dispersion in the fBZ in Figs. S1 and S3 in the SM[46], respectively.

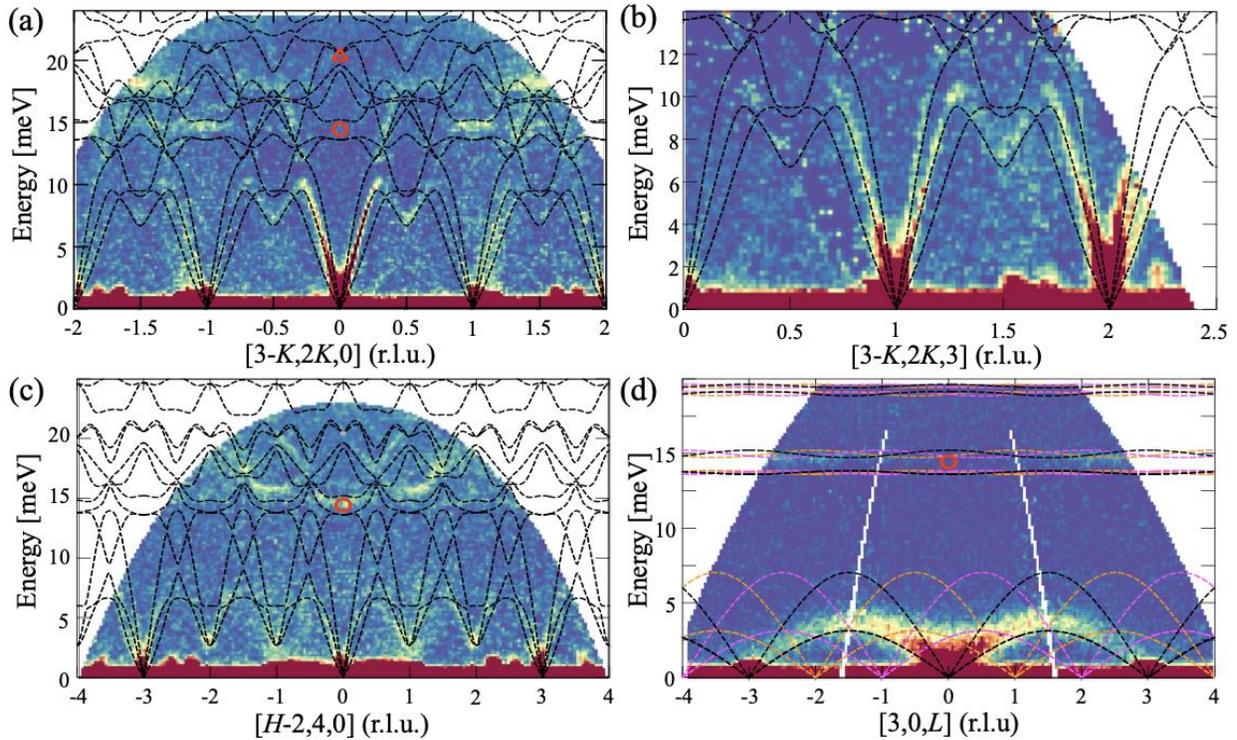

**Figure 3 | Phonon dispersions along high-symmetry paths.** INS measured dynamical structure factors (heat maps) are compared with DFT dispersions using PTS in conventional cells, Eq. 2 (dashed curves). Measured Raman-active phonon modes from Ref. 30, $E_g$ (red circle) and $A_{1g}$ (red triangle), are also shown. Black, pink, and orange dashed curves correspond to integers $l=0$, $l=1$, and $l=2$, respectively. Only black dashed curves are shown in (a-c) for a clear visual comparison.

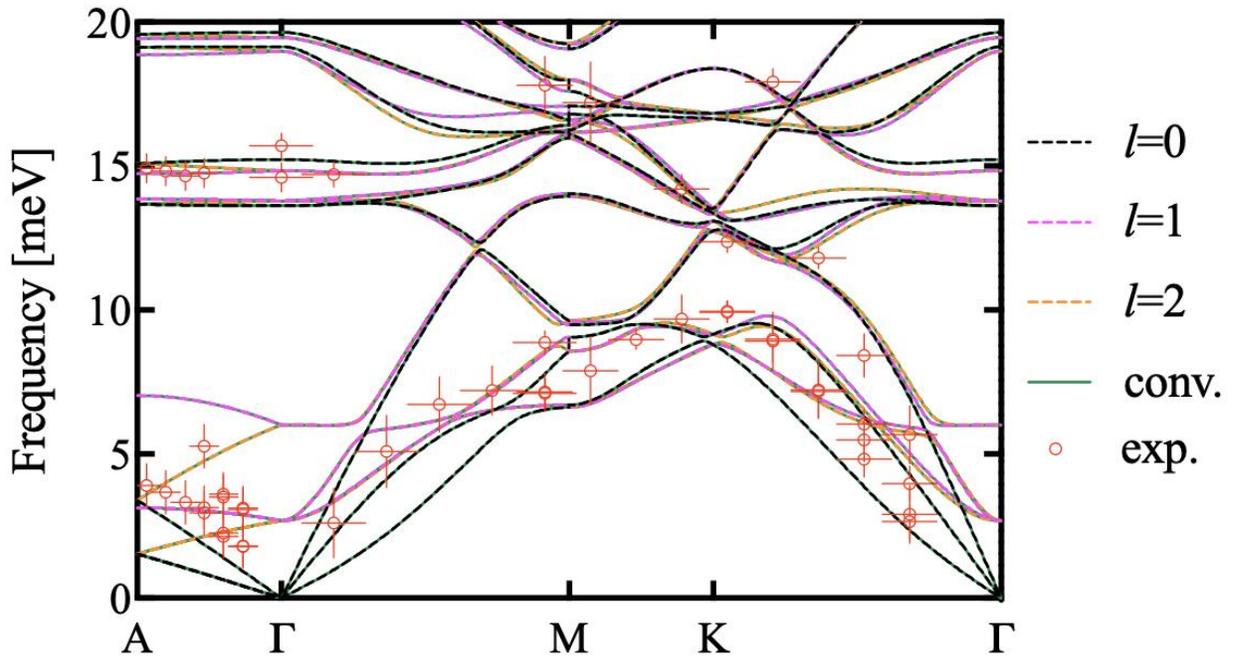

**Figure 4 | Calculated phonon dispersions and extracted measured data points.** The dashed colored curves, solid green curves, and red circles represent dispersions using PTS (Eq. 2), conventional dynamics (Eq. 1), and measurements, respectively. Vertical error bars indicate the full width at half maxima at the measured peak positions (circles), while the horizontal error bars indicate the range of momentum integration of the data. Black, pink, and orange dashed curves correspond to integers $l=0$, $l=1$, and $l=2$, respectively. The dispersion corresponds to calculation #1 in Table 2.

### 4.2 Varying spin configurations and Hubbard corrections

The vibrational behavior of bulk $CrCl_3$ is governed by the underlying atomic structure and magnetic spin configurations of its lattice. Thus, it is natural to ask: to what extent do varying spin configurations and anisotropic couplings govern the phonons, particularly given the varying spin geometries (FM, AFM, PM) attainable in experiments at different temperatures and applied magnetic fields?

As mentioned previously, bulk $CrCl_3$ is a paramagnet above $T_N=14$ K and below this it has in-plane FM and interlayer AFM spin ordering that can be easily polarized into a fully FM state with small applied in-plane magnetic fields[26,29]. Our INS measurements sampled CrCl$_3$ phonons at 30 K, well into the PM state. Given these varying magnetic geometries, developing insights into the sensitivity of the phonons to varying spin configurations and Hubbard corrections is of scientific and practical importance. Figure 5c gives calculated low frequency dispersions for FM and AFM configurations along high-symmetry lines in the fBZ shown in Fig. 5b. The calculations used a Hubbard correction ($U_{eff}$=3.4 eV) for the Cr 3-$d$ electrons and the DFT-D3 functional to describe the weak vdW interplanar interactions. To enable direct comparison, the FM configuration was calculated with a six-layer unit cell. Both in-plane and out-of-plane dispersions generally overlap, suggesting that interlayer spin configuration has negligible effect on the phonon dispersion of bulk CrCl$_3$. This may be attributed to the weak coupling between the layers and relatively large interlayer distances between spins. NM calculations (with $U_{eff}$=3.4 eV and the DFT-D3 functional) give smaller lattice parameters relative to the FM and AFM configurations (Table 1) and, more importantly, strongly imaginary phonons. This suggests that in-plane magnetism is required for dynamical stability in bulk CrCl$_3$. We also tested the relevance of the in-plane spin configurations by comparing phonons with an in-plane AFM spin configuration with the FM case (Fig. S4 in the SM). Surprisingly, like the FM and AFM interlayer cases, there is little variation in the phonon dispersions. These calculations suggest that the spin configurations do not play a significant role in determining the structure and phonons of CrCl$_3$, yet the magnetic interactions are nonetheless required.

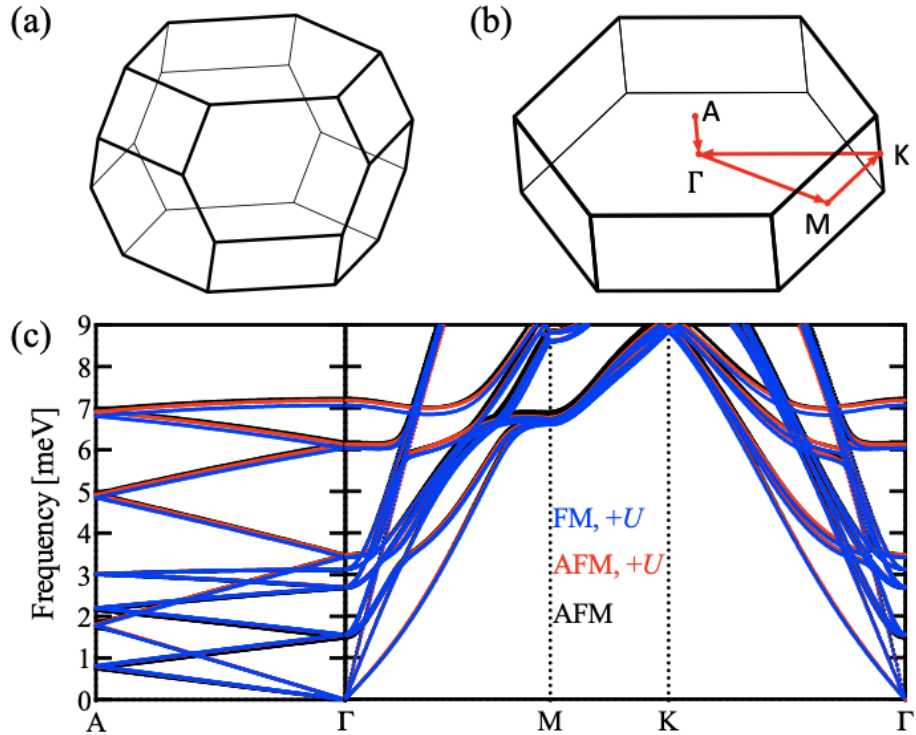

**Figure 5 | Calculated phonon dispersions for FM and AFM configurations and Hubbard corrections.** First Brillouin zones of $CrCl_3$ are shown for (a) primitive and (b) conventional unit cells. All phonon calculations presented here correspond to the conventional cell Brillouin zone. (c) The phonon dispersion for the FM configuration was calculated with a six-layer unit cell. The high-symmetry paths are shown in (b). The phonon dispersion for the AFM configuration without Hubbard corrections is also included. All calculations include the DFT-D3 functional for weak vdW interactions.

Hubbard corrections via the DFT+$U$ method have been included to improve the description of the 3-$d$ electrons of the Cr atoms, which significantly change the electronic band structure as demonstrated in Fig. 6 for the AFM configuration with and without the Hubbard $U$ correction. The energy was set to zero for the highest occupied band in each case. The calculated projected density of states (PDOS) of $d$-orbitals of Cr atoms overlaps with the total DOS in the energy range around the Fermi energy demonstrating that the Cr $d$-orbitals are indeed responsible for variation with $U$. We note that the band gap increases from 1.65 to 2.17 eV when Hubbard corrections are included, closer to the result of 3.1 eV measured in a previous study[26]. Unlike the electronic states, the phonon dispersion of the AFM configuration does not vary with and without the Hubbard corrections as shown in Fig. 4c.

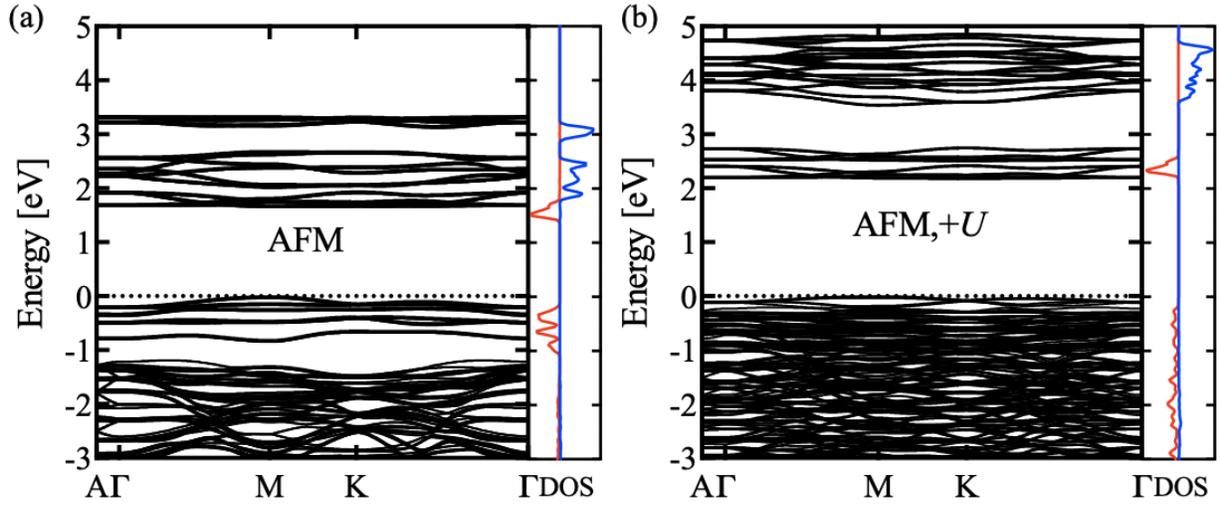

**Figure 6 | Calculated electronic band structures for the AFM spin configuration.** (a) without Hubbard corrections and (b) with $U$=3.4 eV. PDOS of $d$-orbitals of Cr are shown by blue (spin-up) and red (spin-down) curves. All calculations include the DFT-D3 functional for vdW interactions.

**4.3 vdW interactions and compatibility with SOC**

Bulk CrCl$_3$ is a layered material with weak cross-plane vdW interactions compared with covalent in-plane bonding. Here, we investigate the effect of varying vdW functionals and SOC in the calculation of the structure and phonon dispersions.

*Structure* – vdW interactions are critical to describing weak layer coupling in a variety of materials[35,38] – the cleavable $X$Cl$_3$ magnets are no exception. For most FM and AFM calculation configurations (see **Table 1**) with varying $U$ and SOC, including vdW interactions (DFT-D3 and vdW-DF-optB86b functionals) gives relaxed lattice parameters ($a$ = 5.989-6.052 Å; $c$ = 17.244-17.394 Å) in reasonable agreement with measured values ($a$ = 5.942 Å; $c$ = 17.333 Å)[40]. Relaxation of the lattice without vdW interactions give unreasonable structures with $c$ > 19 Å. Attempts to relax the lattice with vdW-DF-optB8b and SOC yielded significantly erroneous results. Analysis of the output file suggested that introduction of SOC resulted in a large perturbation to the vdW-DF-optB86b functional. A potential cause may be the lack of a spin-component on the non-local correlation (vdW-DF) as implemented in VASP, which invalidates calculations that include both SOC and vdW-DF[49].

*Phonon dispersions* – vdW interactions and structural differences can translate into variations in the phonon dispersions, particularly for the cross-plane direction. Fig. 7 gives the low frequency phonon dispersions for the in-plane and cross-plane directions of bulk $CrCl_3$ in the FM spin configuration for three cases: (1) DFT-D3 without SOC, (2) vdW-DF-optB86b without SOC, and (3) DFT-D3 with SOC. Similar to the structural parameters, incorporating SOC has very little effect on the dispersion of $CrCl_3$ (green and red curves). Similarly, SOC was previously found to have little to no influence on calculated phonons of $CrBr_3$ and $CrI_3$ monolayers[50]. The vdW-DF-optB86b functional gives slightly higher phonon frequencies than those given by the DFT-D3 functional, likely a consequence of vdW-DF-optB86b giving a *c* lattice parameter that is 0.9% smaller.

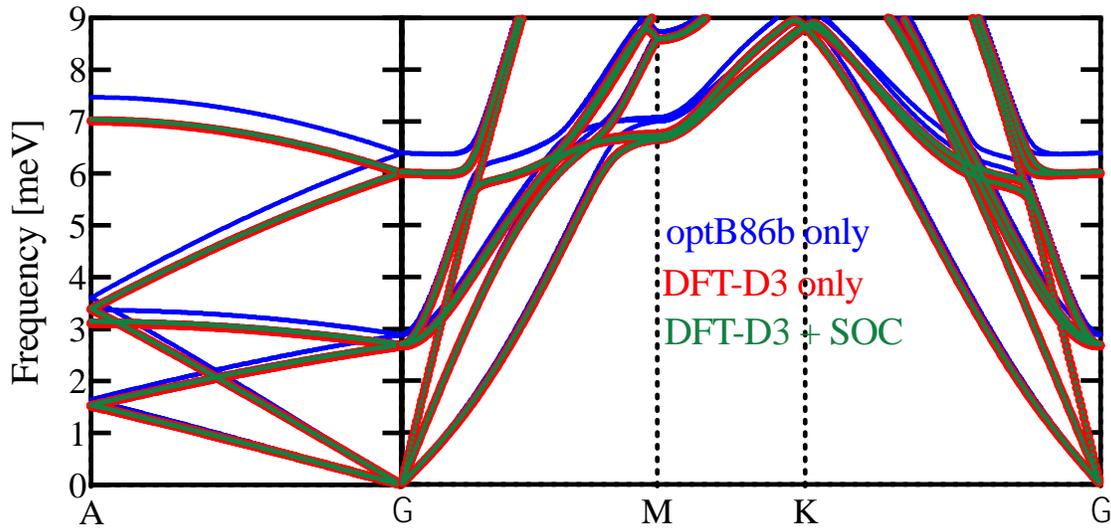

**Figure 7 | Calculated dispersions for FM $CrCl_3$ with varying vdW and SOC interactions.** The red curves overlap with green curves, suggesting negligible effect of SOC.

## 5. Discussion

The transport of phonons with coupling to spins is key to understanding quantum thermal Hall effects[16,51-54] and anomalous magnetic-field-induced thermal conductivity behaviors in α-$RuCl_3$[17] and $CrCl_3$[13]. Below room temperature, magnetic excitations are the main phonon scatterers, which can be tuned and eliminated by in-plane magnetic field. As a result, measured thermal conductivity values increase sharply in high field and become much larger than the zero-field values[13,17]. Indeed, our calculations demonstrate that magnetism is critical to determining the structure and vibrations of $CrCl_3$, without which calculations suggest strong dynamical instability

(large magnitude imaginary phonons). However, the CrCl$_3$ structure and phonon dispersions are not sensitive to the static spin configuration, giving similar properties for different AFM and FM systems (Figs 5 and S4). This suggests that phonons coupling to spin-excitations in the field-dependent transport experiments on CrCl$_3$[13] are indeed a dynamical effect and not related to static changes in the local spin configurations.

Not surprisingly, first principles-based calculations demonstrate that vdW interactions are necessary to describe the structures and phonons for CrCl$_3$, while SOC is not. SOC is expected to be relatively weak given the small atomic number of the constituent Cr and Cl atoms. Furthermore, the t$_{2g}$ bands are half filled for octahedrally bonded Cr$^{3+}$ and the e$_g$ bands are significantly separated from the t$_{2g}$ bands due to crystal field splitting. In this case, the orbital moment is negligible, and thus SOC is also. It is also found that Cr ions contribute negatively to the first order SOC in a similar system CrI$_3$[55]. As a result, the effect of SOC on lattice thermal conductivity in CrCl$_3$ is likely small.

Exploiting the PTS within conventional cells provides an efficient way to study dynamics in a conventional cell at a similar numerical cost to using a primitive cell and more in line with typical experimental geometries. As shown in Fig. 5a, the fBZ of the primitive unit cell for CrCl$_3$ is rhombohedral with 14 faces formed at seemingly arbitrary angles. This brings difficulty in evaluating the symmetry in reciprocal space for phonon calculations, particularly when interested in cross-layer and in-layer directions. Using a conventional cell, however, results in a hexagonal fBZ that is more symmetric (see Fig. 5b), though at the cost of a larger number of atoms. Calculations based on the conventional cell using PTS are cheaper than those for conventional dynamics, which mitigates the disadvantage of using conventional cells.

Note that using PTS to simplify conventional dynamics is not limited to phonons in CrCl$_3$, but can be generally applied to different quasiparticles (*e.g.*, Bloch electrons, magnons) in any material (vdW layered or covalently bonded) with translational symmetry in the conventional cell. This method will be useful for studying phonons and thermal transport in large complex materials.

## 6. Summary

In summary, our investigations on bulk CrCl$_3$ phonons suggest that in-plane spins are essential but specific spin configurations and SOC do not matter, which provides insights into recent magnetic-field-dependent thermal transport measurements. We also introduce an efficient method that uses primitive translational symmetry within conventional cells to give insights into the underlying phase dynamics of phonon dispersions built from conventional cells and elucidates quantum phase interference conditions that assist in the interpretation of measured INS spectra. Our work advances the understanding of phonons in conventional cells of various materials and their dependence on spin configurations in layered magnets.

**Acknowledgements**

We thank Simon Thébaud and Li Yin for useful discussions. This work was supported by the US Department of Energy, Office of Science, Office of Basic Energy Sciences, Material Sciences and Engineering Division. The calculations used resources of the Compute and Data Environment for Science (CADES) at the Oak Ridge National Laboratory, which is supported by the Office of Science of the U.S. Department of Energy under Contract No. DE-AC05-00OR22725.

**Author contributions**

X. L. performed the lattice dynamics and first-principles calculations. S. D., G. G. and A. C. carried out INS measurements. J. Y. and M. M. grew the crystal. X. L. and L. L. analyzed the data and constructed the paper with contributions from S. M., T. B., and V. C.. L. L. supervised the research. All authors commented on, discussed, and edited the paper.

**Data availability**

The data that support the findings of this study are available from the corresponding authors on reasonable request.

**Declaration of competing interests**

The authors declare no Competing Financial or Non-Financial Interests.

# Supplemental Materials

**Phonons and phase symmetries in bulk CrCl$_3$ from scattering measurements and theory**

X. Li[1], S.-H. Do[1], J.-Q. Yan[1], M. A. McGuire[1], G. E. Granroth[2], S. Mu[3], T. Berlijn[4,5], V. R. Cooper[1], A. D. Christianson[1*], and L. Lindsay[1†]

[1]Materials Science and Technology Division, Oak Ridge National Laboratory, Oak Ridge, Tennessee 37831, USA

[2]Neutron Scattering Division, Oak Ridge National Laboratory, Oak Ridge, Tennessee 37831, USA

[3]Materials Department, University of California, Santa Barbara, California 93106, USA

[4]Center for Nanophase Materials Sciences, Oak Ridge National Laboratory, Oak Ridge, Tennessee 37831, USA

[5]Computational Sciences and Engineering Division, Oak Ridge National Laboratory, Oak Ridge, Tennessee 37831, USA

Corresponding email: christiansad@ornl.gov* (experiment); lindsaylr@ornl.gov[†] (theory)

This file includes:
1. Derivation of Eq. 3 in the main text
2. Numerical details for DFT calculations
3. Supplemental figures for measured and calculated phonon dispersions

# 1. Derivation of Eq. 3

The dynamical structure factor for the conventional geometry is[43,44]

$$S(\mathbf{Q}, E) = \frac{1}{2N_{uc}} \sum_{\mathbf{q},J,\mathbf{G}} \left| \sum_K \frac{b_K}{\sqrt{m_K}} e^{-i\mathbf{Q}\cdot\mathbf{r}_K} (\mathbf{Q} \cdot \boldsymbol{\xi}_{\mathbf{q},J,K}) e^{-W_K} \right|^2 \frac{(n_{\mathbf{q},J} + 1)}{\omega_{\mathbf{q},J}} \delta(E - \omega_{\mathbf{q},J}) \delta(\mathbf{Q} - \mathbf{q} - \mathbf{G}) \quad \text{(S1)}$$

where $\mathbf{Q}$ and $E$ are the momentum and energy transfer, respectively, $N_{uc}$ is the number of conventional unit cells in the crystal, $\mathbf{G}$ is a reciprocal lattice vector, $b_K$ is the average nuclear scattering length of atom $K$, $m_K$ is its mass, $\mathbf{r}_K$ is the position of atom $K$ in the conventional unit cell, $\boldsymbol{\xi}$ is its conventional eigenvector, $W_K$ is the temperature-dependent Debye-Waller factor, $n_{\mathbf{q},J}$ and $\omega_{\mathbf{q},J}$ are the equilibrium Bose-Einstein distribution and frequency, respectively, of a phonon mode with wavevector $\mathbf{q}$ and polarization $J$.

Using an extended zone beyond the first Brillouin zone, the reciprocal lattice vector $\mathbf{G}$ can be ignored. Applying the momentum ($\mathbf{Q} = \mathbf{q}$) and energy ($E = \omega_{\mathbf{q},J}$) delta functions, we have

$$S(\mathbf{Q}, E) = \frac{1}{2N_{uc}} \sum_J \left| \sum_K \frac{b_K}{\sqrt{m_K}} e^{-i\mathbf{Q}\cdot\mathbf{r}_K} (\mathbf{Q} \cdot \boldsymbol{\xi}_{\mathbf{Q},J,K}) e^{-W_K} \right|^2 \frac{(n_{\mathbf{Q},J} + 1)}{\omega_{\mathbf{Q},J}} \quad \text{(S2)}$$

Using primitive translational symmetry (PTS), atom $K$ in the conventional unit cell is represented by atom $k$ in layer $h$. The eigenvector can be expressed as

$$\boldsymbol{\xi}_{\mathbf{q},J,K} = \frac{1}{\sqrt{N}} \boldsymbol{\varepsilon}_{\mathbf{q},j,l,k} e^{ih\mathbf{q}\cdot\mathbf{S}} e^{i2\pi l h/N} \quad \text{(S3)}$$

where the polarizations $J$ are now labeled by $l$ and $j$, $\mathbf{S}$ is a translation vector that maps layers within the conventional unit cell using PTS, $N$ is the number of layers, and $\boldsymbol{\varepsilon}$ is the eigenvector determined from Eq. 2 of the main manuscript (using PTS). In this notation, the position of atom $K$ is

$$r_{hk} = h\mathbf{S} + \Delta_k \quad \text{(S4)}$$

where $\Delta_k$ locates the atom with respect to layer $h$. Using Eqs. S3 and S4 in Eq. S2, the dynamical structure factor is

$$S(\mathbf{Q},E) = \frac{1}{2N_{\text{uc}}N}\sum_{jl}\left|\sum_{h,k}\frac{b_k}{\sqrt{m_k}}e^{-i\mathbf{Q}\cdot(h\mathbf{S}+\Delta_k)}(\mathbf{Q}\cdot\boldsymbol{\varepsilon}_{\mathbf{Q},j,l,k}e^{ih\mathbf{Q}\cdot\mathbf{S}}e^{i2\pi lh/N})e^{-W_{h,k}}\right|^2\frac{(n_{\mathbf{Q},j}+1)}{\omega_{\mathbf{Q},j}} \quad (S5)$$

Rearranging terms gives

$$S(\mathbf{Q},E) = \frac{1}{2N_{\text{uc}}N}\sum_{jl}\left|\sum_{h=0}^{N-1}e^{i2\pi lh/N}\sum_{k}\frac{b_k}{\sqrt{m_k}}e^{-i\mathbf{Q}\cdot\Delta_k}(\mathbf{Q}\cdot\boldsymbol{\varepsilon}_{\mathbf{Q},j,l,k})e^{-W_{h,k}}\right|^2\frac{(n_{\mathbf{Q},j}+1)}{\omega_{\mathbf{Q},j}} \quad (S6)$$

which is Eq. (3) in the main manuscript.

## 2. Numerical details for DFT calculations

DFT calculations of the electronic structure of bulk CrCl3 were based on the projector augmented wave method (PAW) as implemented in the Vienna Ab-initio Simulation Package (VASP)[56-59]. The generalized gradient approximation, parameterized by Perdew, Burke, and Ernzerhof (PBE)[60] was used for exchange-correlations. A 520 eV kinetic energy cutoff in the plane-wave expansion and energy convergence criteria of 10-6 eV were employed. Ionic relaxations were performed until Hellmann-Feynman forces converged to 10-4 eV/Å. The $R\bar{3}$ structure was optimized with a Γ-centered 4x4x4 k-mesh. The harmonic IFCs were calculated using the finite displacement method implemented in the phonopy package[61] with Γ-centered 2x2x2 k-meshes.

## 3. Supplemental figures

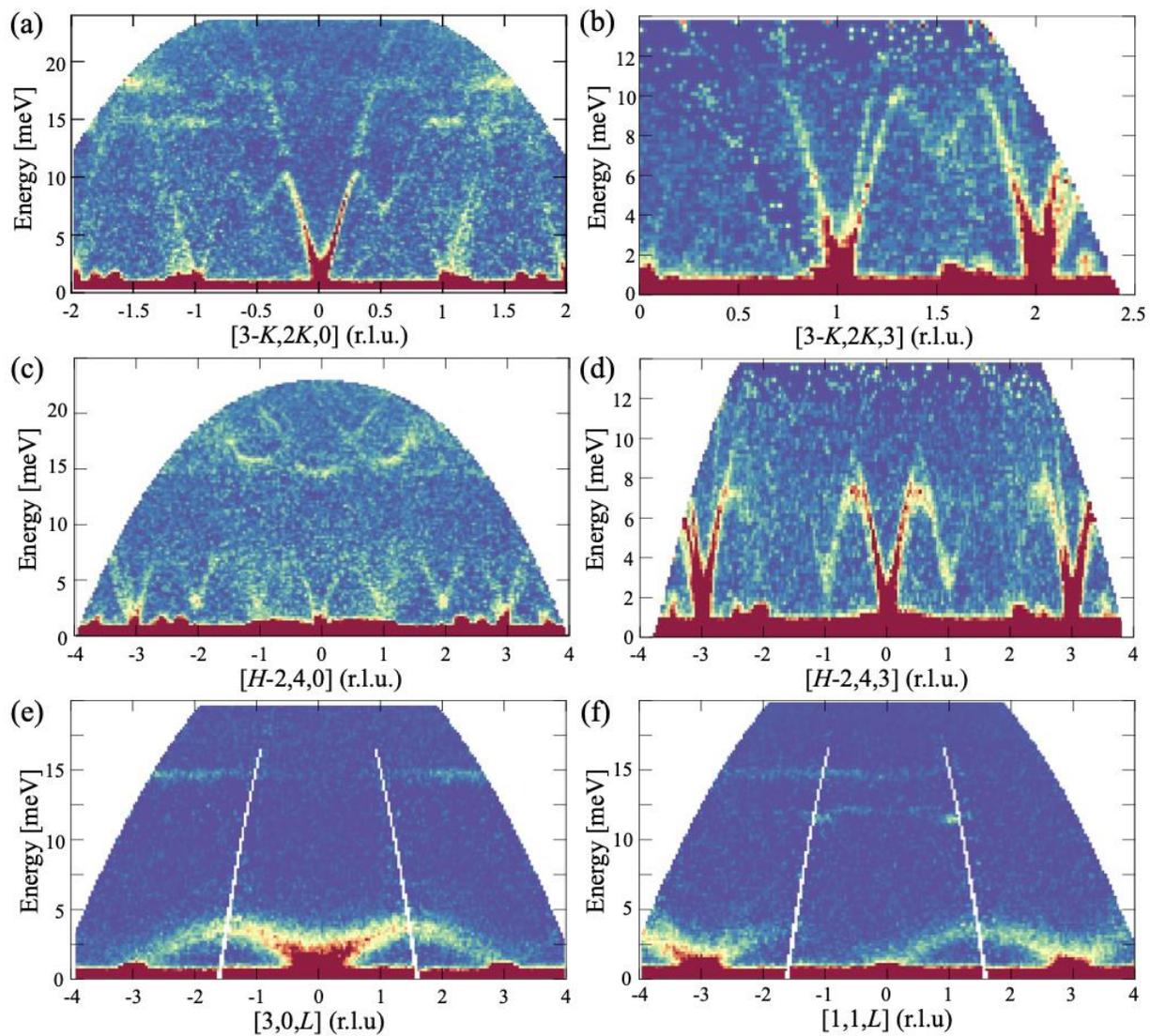

**Figure S1 | Observed phonon spectra of bulk CrCl$_3$ with INS along high-symmetry directions, measured at 30 K.**

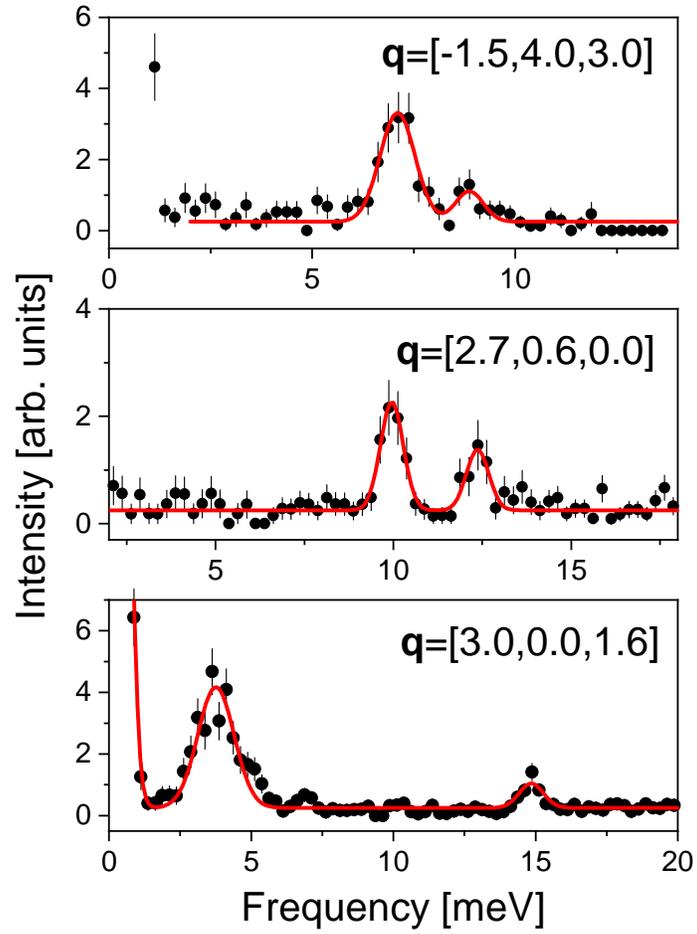

**Figure S2 | Constant momentum scan in the high-symmetry direction.** The red solid lines indicate the Gaussian fittings to the excitation peaks of the scan (black circles).

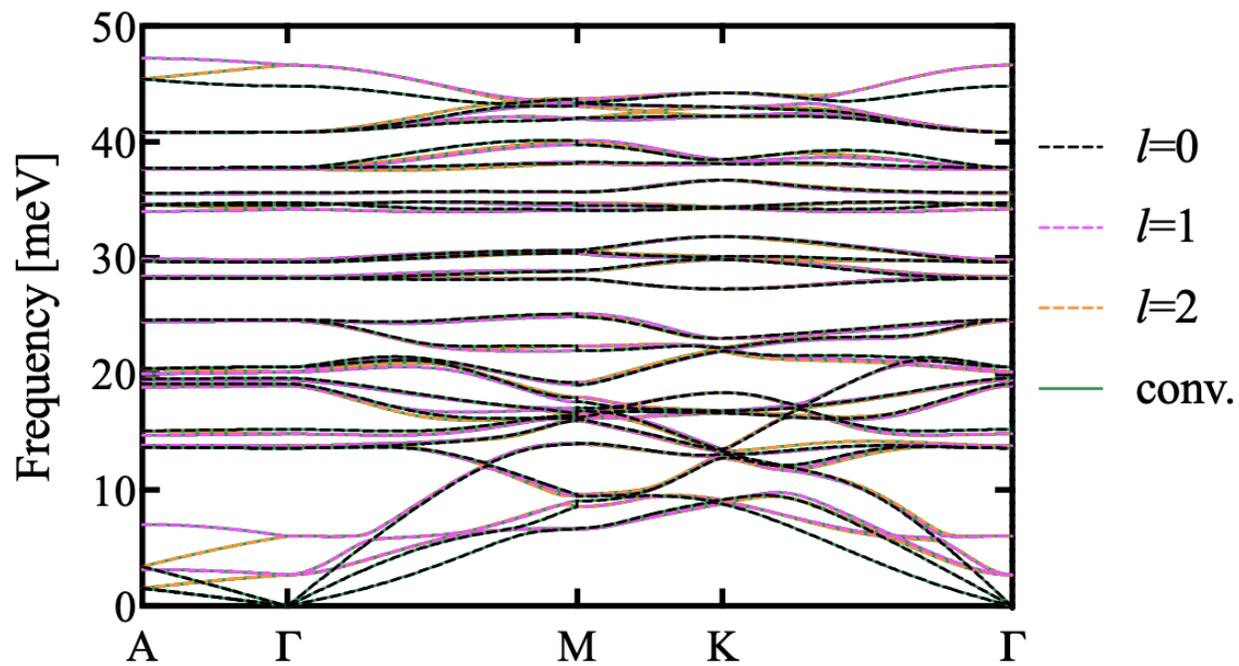

**Figure S3 | Calculated full phonon dispersion of bulk CrCl$_3$ along high-symmetry directions in the first Brillouin zone.** The calculations and legends are the same as in Figure 4 in the main text.

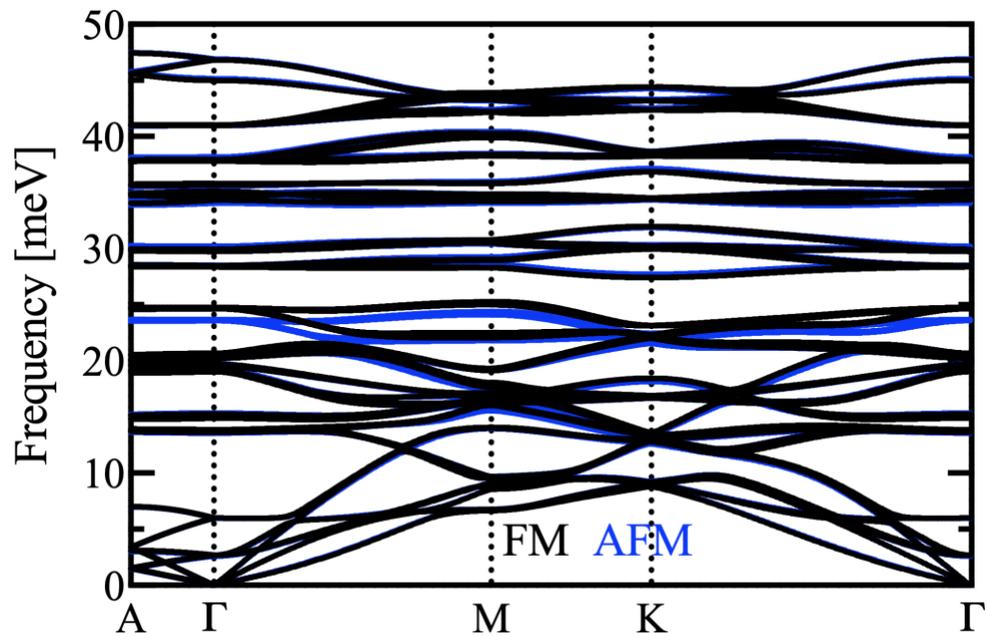

**Figure S4 | Calculated phonon dispersion with different in-plane spin configurations.** There is little variation in the dispersion between the in-plane FM (black curves) and in-plane AFM (blue curves) configurations. The structure is fully relaxed in each case (see Table 2 in main text).